\documentclass[twoside,twocolumn]{revtex4}
\begin{document}
%\frontmatter

\title{Instantons and the $\theta$ vacuum in non-linear $\sigma$ model and QCD}

\author{H.~Albert\vspace{0.2cm}}
%\date{}

\affiliation{\text{II}.Institut f\"ur Theoretische Physik, Universit\"at Hamburg,\\
       Luruper Chaussee 149, 22761 Hamburg, Germany}
%\affiliation{Hamburg University, Germany}

%%%%%%%%%%%%%%%%%%%%%%%%%%%%%%%%%%%%%%%%%%%%%%%%%%%%%%%%%%%%%%
%
% Abstract
%
%%%%%%%%%%%%%%%%%%%%%%%%%%%%%%%%%%%%%%%%%%%%%%%%%%%%%%%%%%%%%%
\begin{abstract}
This is a talk given on the $\theta$ vacuum and cluster properties
in QCD and the non-linear sigma model, proposed to be given on
occasion of an invitation to the seventh international conference on
Symmetry in Nonlinear Physics, June 24 - 30, 2007, Kiev.
\end{abstract}
\maketitle
%%%%%%%%%%%%%%%%%%%%%%%%%%%%%%%%%%%%%%%%%%%%%%%%%%%%%%%%%%%%%%
\section{Introduction}
%%%%%%%%%%%%%%%%%%%%%%%%%%%%%%%%%%%%%%%%%%%%%%%%%%%%%%%%%%%%%%
The non-linear sigma model in two dimensions [1,2] is a field theoretical description of the Heisenberg ferromagnet. For sufficiently low temperature the interaction between the local magnets becomes dominant and the local magnets tend
to align. Thus we get an ordered state characterized by an order parameter, which can be chosen to be the direction of the local spin vector. Thus the order parameter is a unit vector, which can be represented by a triple of scalar fields $[\varphi^1 (x) ,\varphi^2 (x) ,\varphi^3 (x)]$, subject to the constraint $\varphi^b \varphi^b = 1 $.
The static energy functional of the Heisenberg ferromagnet is given by
$$H[\varphi ] = \langle d\varphi^b | d\varphi^b \rangle = \int \partial_j \varphi^b \partial^j \varphi^b \sqrt{g} d^D x$$
with
$$\varphi^b \varphi^b = 1\text{.}$$
Since we want to classify all finite energy solutions of the non-linear sigma model, we have to impose $$H[\varphi ] = \langle d\varphi^b | d\varphi^b \rangle  < \infty$$which leads to the boundary condition$$\lim_{x\to\infty} d\varphi^b (x) = 0$$
and that $\varphi^b$ must be asymptotically constant [11]
$$\lim_{x\to\infty} \varphi^b (x) = \varphi_0^b\text{.}$$
Which components the constant vector at infinity has cannot be said.
All unit vectors are equally likely.
Without restriction of generality we follow convention and choose the north pole as the asymptotic value
$$\lim_{x\to\infty} \varphi^b = [0;0;1]\text{.}$$
From the very beginning we chose the vector $\varphi^b $ to be a triple. But nothing prevents us from  asking whether we can extend the model with $\varphi^b $ being an N-tuple. But in the next section we will see with the help of homotopy theory that for N $>$ 3 there are no nontrivial vacuum solutions. Finally it should be said that the global (non space-time dependent) symmetry group of the non-linear sigma model is O(N) for $\varphi^b$ being a N-tuple or especially O(3)
for the Heisenberg ferromagnet, i.e. $\varphi$ being a triple.
As a final comment we should say that we will not discuss generalized non-linear sigma models,
where space-time and target (field)-space may have arbitrary geometries [7].
%
%
%%%%%%%%%%%%%%%%%%%%%%%%%%%%%%%%%%%%%%%%%%%%%%%%%%%%%%%%%%%%%%
\section{Instantons and the non-linear sigma model }
%%%%%%%%%%%%%%%%%%%%%%%%%%%%%%%%%%%%%%%%%%%%%%%%%%%%%%%%%%%%%%
In this section I will make use of the exterior derivative calculus, since use is made of
the duality operation, which can be written best in these terms.
Due to the constraint $\varphi^2= 1$, the scalar field in the non-linear sigma model
(or Heisenberg ferromagnet) defines a mapping
$$\varphi^a :R^2 \rightarrow S^{N-1}\text{.}$$
We can compactify $R^2$ by making use of the boundary
condition (see above in the introduction) on $\varphi^a$:
$$\lim_{x\to\infty} \varphi^a (x) = (1, 0, 0..., 0)$$
All points at infinity on $R^2$ can be identified, giving a map $f: R^2 \rightarrow S^2$
lifting the map $\varphi^a$ to the map $\varphi^a: S^2 \rightarrow S^{N-1}$.
But that means, that $\varphi^a$ is a representative of an element of $\pi_2 (S^{N-1} )$.
Now it is wellknown that $\pi_2 (S^{N-1} ) = 0$ for N equal or larger than 4.
This can best be seen by using fiber bundle techniques:
Define a principal fiber bundle:
$$O (N) \stackrel{\pi}{\longrightarrow} \frac {O (N)}{O (N-1)}\text{,}$$ with fiber O (N-1). Now
O (N) is the translational group on the sphere $S^{N-1}$, while O (N-1) leaves a point on the sphere fixed.
Hence, the quotient space $O (N)/O (N-1)$ is isomorphic to $S^{N-1}$.
We then have an exact sequence of homotopy groups:
$$...\longrightarrow \pi_i (O (N-1)) \stackrel {\alpha}{\longrightarrow} \pi_i (O(N) ) \longrightarrow \pi_i (S^{N-1} )$$
                                                      $$  \longrightarrow \pi_{i-1} (O(N-1)) \stackrel {\beta}{\longrightarrow} ...$$
where $\alpha$ is an epimorphism (surjective), $\beta$ is an
isomorphism for $i \le N-2$. Since the sequence is exact, it
follows, that $\pi_i (S^{N-1}) = 0$ for $i \le N-2$ (see [5,6]).
Especialy we have $$SO(3) \longrightarrow SO(4) \longrightarrow
S^3$$ From the homotopy sequence we have$$0 = \pi_2 (SO(3))
\longrightarrow
\pi_2 (SO(4)) \longrightarrow \pi_2 (S^2 ) = 0$$ Therefore we have $0 =
\pi_2 (SO(n)) = \pi_2 (O(n)) $ for $n \le 4$ since $\pi_2 (SO(3)) =
\pi_2 (RP^3 )) = \pi_2 (S^3 )) = 0$
In the following we restrict $\varphi$ to be a triple, and
hence the symmetry group of the sigma model being O(3). It is a
representative of an element of $\pi_2 (S^2)$ ($\pi_2$ else being
zero, as explained above). The space of solutions is divided into
sectors, labeled by n. n meaning the mapping degree or winding
number. The Hamiltonian of the sigma- model can be cast into a
form, exhibiting directly the splitting of the space of solutions
into sectors, labeled by n:
$$H[\varphi] = \frac{1}{4} ||*d\varphi^a \mp \# d\varphi^a ||^2 \pm 4\pi
n(\varphi )\text{,}$$ where the $*$ operation means the Hodge dual of the one form and the
operation $\#$ is defined in the section ``Duality condition''.
The decomposition of the Hamiltonian runs as follows: For it we need two more concepts:\begin{itemize} \item{The selfduality equation, including conformal invariance} and \item{an analytical expression for the winding number or, what is the same, the topological charge.}\end{itemize}
\section{Duality condition}
The derivatives $\partial_i \varphi$ are tangent vectors to the sphere $S^2$ in field space, made
up by the fields $\varphi^a$ due to the constraint $\varphi^2 = 1$. We introduce a duality
operation, which will be called $\#$, by the formula
$$\# d\varphi^a =\epsilon_{abc} \varphi^b d\varphi^c\text{.}$$
This corresponds to a rotation by $\frac{\pi}{2}$ in the tangent space to the sphere $S^2$ in field
space, described above.
%%%%%%%%%%%%%%%%%%%%%%%%%%%%%%%%%%%%%%%%%%%%%%%
\section{winding number}
The winding number tells us, how often a topological space (not necessarily $S^N$) is mapped on another
topological space. The winding number is stable under continuous deformation of the mapping. The crucial point is that the mappings can not always be deformed to the constant mapping.
%%%%%%%%%%%%%%%%%%%%%%%%%%%%%%%%%%%%%%%%%%%%%%%
\section{The winding number or The Brouwer theorem}
To introduce the Brouwer index, we need the concept of the degree of a map (smooth):
\\{\bf Definition}:
\\The degree of a smooth map $f: \cal M \longrightarrow \cal N$, $\cal M , \cal N$ being compact, orientable manifolds of the same dimension, at a regular value $Q \epsilon \cal N$
 is the integer $$Deg (f;Q) = \sum_{P_i \epsilon f^{-1} (Q)} sgn | \partial y^i /\partial x^j |_{P_i}\text{,}$$
 where $sgn |\partial y^i /\partial x^j | = \pm  1$ according to whether $f^*$ preserves or reverses orientation.
 \\{\bf Brouwer's theorem}:
\\ Let $f: \cal M \longrightarrow \cal N$ be be a smooth map, T an n-form on $\cal N$. Then:
 $$\int_{\cal M} f^* T = Deg (f) \int_{\cal N} T\text{.}$$
 Now lets choose T as $\epsilon = (g)^{1/2} dx^1\wedge ... \wedge dx^n$, the volume form on $\cal N$. Then we get
 $$\int_{\cal M} f^* \epsilon = Deg(f)\int_{\cal N} \epsilon = Deg(f) Vol[\cal N]$$ or
 $$Deg(f) = \frac{\int_{\cal M} f^* \epsilon}{Vol[\cal N]}$$
 The map in our $\sigma$ model is given by: $\varphi^a : R^2 \longrightarrow S^{N-1}$, $\varphi^2 = 1$
 and the winding number by
 $$Deg(\varphi) = \frac{1}{Vol(S^{N-1})} \int_{R^2} \epsilon_{a_0...a_n}
 \varphi^{a_0} d\varphi^{a_1} ...d\varphi^{a_n}\text{.}$$
 We know now, that $\pi_2 (O(N))$ is zero for N larger than 3, so we restrict to the case N = 3.
 $$\Rightarrow Deg(\varphi ) = n = \frac{1}{8\pi} \int_{S^2} \epsilon_{abc} \varphi^a d\varphi^b \wedge
 d\varphi^c\text{,}$$
 in coordinates $n = \frac{1}{4\pi} \int \epsilon_{abc} \epsilon_{ij} \varphi^a \partial_i \varphi^b \partial_j \varphi^c dx^1 dx^2 $,
 or
 $$n = \frac{1}{8\pi} <*d\varphi^a |\# d\varphi^a>\text{.}$$
 Finally, we observe $$<d\varphi^a | d\varphi^a> = <*d\varphi^a |*d\varphi^a> = <\# d\varphi | \#
 d\varphi^a>\text{,}$$
 since as already said above a
 duality transformation in two dimensions amounts to a rotation about $\frac{\pi}{2}$,
 which leaves the norm ($||\;\;||^2 =\langle | \rangle$)
 invariant.
 Now we have all the ingredients, to invoke the Bogomolny decomposition.
 $$H = \frac{1}{2} <d\varphi^a | d\varphi^a>$$ $$= \frac{1}{4} <*d\varphi^a | *d\varphi^a > + \frac{1}{4} <\# d\varphi^a |\# d\varphi^a> +$$ $$ \frac{1}{2} <*d\varphi^a |\# d\varphi^a> - \frac{1}{2} <*d\varphi^a | \# d\varphi^a>$$
$$= \frac{1}{2} <*d\varphi^a - \# d\varphi^a |*d\varphi^a - \# d\varphi^a> + 4\pi n(\varphi)$$
$$= \frac{1}{2} ||*d\varphi^a - \# d\varphi^a ||^2 + 4\pi n(\varphi)$$
This form of the Hamiltonian is called the Bogomolny decomposition.
From this formula follows\begin{itemize}
\item{ The energy is bounded below by $4\pi n(\varphi)$}
\item {A configuration with winding number n is a ground state if and only if it satisfies the first order
   differential equation $$*d\varphi^a = \# d\varphi^a$$ if n is positive and
   $$*d\varphi = -\# d\varphi^a$$ if n is negative.}\end{itemize}
   Ground state solutions in the sigma model are called spin waves.
   Pictorially the winding number (or topological charge) describes, how often the spins, aligned along the, say, x-axis, twist
   around this axis, the whole chain of spins being held fixed at both "ends" by boundary conditions.
   These equations are so called double self-dual equations and play a decisive role in classifying the solutions.
The next step will be the proof of the following\\ {\bf Proposition}:
\\A spin configuration $\varphi^a : R^2 \longrightarrow S^2$ is a spin wave if and only if it is a conformal
map from the plane to the sphere.
\\Proof:
\\Let $e_1, e_2$ be a canonical basis of $R^2$. These are lifted to the tangent vectors $\partial_1 \varphi^a , \partial_2 \varphi^a$.
If $\varphi^a$ is a conformal map, this forces $\partial_1 \varphi^a , \partial_2 \varphi^a$ to be
orthogonal and of the same length.
Now we know, that a solution $\varphi^a$ with winding number n is a solution of the following
equations: $$\partial_1 \varphi = \# \partial_2 \varphi , \partial_2 \varphi = -\# \partial_1 \varphi$$
But this shows, that the vectors $\partial_i \varphi^a$ are orthogonal, since the operator
$\#$ amounts to a rotation of $\frac{\pi}{2}$, as already said above. They are also of equal length, since
$\#$ does not alter the length. We now have to show the converse, i.e., that any conformal map
solves the two differential equations above:
Let $\varphi^a : R^2 \longrightarrow S^2$ be a smooth map and introduce the following vectors:
$$P^a = *d\varphi^a - \# d\varphi^a  i.e. P_i = \epsilon_{ij} \partial_j \varphi - \varphi \times \partial_i \varphi$$
$$Q^a = *d\varphi^a + \# d\varphi^a i.e. Q_i = \epsilon_{ij} \partial_j \varphi + \varphi \times \partial_i \varphi$$
Then we get $$P_1 Q_1 = P_2 Q_2 = |\partial_1 \varphi |^2 - |\partial_2 \varphi |^2$$ $$P_1 P_2 = Q_1 Q_2 = 0$$
and
$$P_1 Q_2 = P_2 Q_1  = -2\partial_1 \varphi \partial_2 \varphi\text{.}$$
But for a conformal map $\varphi : R^2 \longrightarrow S^2$, $\partial_1 \varphi , \partial_2 \varphi $ are orthogonal tangent vectors of the same length and hence the right hand sides of the equations above vanish. This shows that the $P_1 , P_2 , Q_1 , Q_2$ are mutually orthogonal.
The next and final step is to show that\\{\bf Lemma}\\ If $\varphi$ is a conformal map the either
$$P_1 = P_2 = 0$$
or
$$Q_1 = Q_2 = 0\text{.}$$
For this we need the following {\bf remark}:
If $\partial_1 \varphi$ (or $\partial_2 \varphi $) vanishes, then all four vectors $P_1, P_2, Q_1, Q_2$ vanish. To
see this, let $\partial_1 \varphi$ vanish. Then we get $$P_2 = - \varphi \times \partial_2 \varphi$$and
         $$Q_2 = \varphi \times \partial_2 \varphi$$ But since they are orthogonal vectors, $\partial_2 \varphi$ has also to vanish.
Clearly $P_1, P_2, Q_1, Q_2$ all have to vanish.
To prove the proposition we observe, that $P_1, P_2, Q_1, Q_2$ all are tangent vectors in the same two-dimensional
plane. Since they are all mutually orthogonal, two of them have to vanish. But we have to exclude the possibilities, that either
two $P_i, Q_j$ vanish, leaving the other two nonzero.
Suppose now, that $P_1, Q_1$ is equal to zero. Then $\partial_2 \varphi =\frac{1}{2} (P_1 + Q_1 )$ vanishes and hence all $P_i , Q_j $ vanish. The same amounts for $P_2 , Q_2 $.  Suppose now that $Q_2 = P_1 = 0$. From $Q_2 = 0$ we get $\partial_1 \varphi = \varphi \times \partial_2 \varphi$ .
Inserting this into the expression for $P_1$, we obtain
$$0 = P_1 = \partial_2 \varphi - \varphi \times (\varphi \times \partial_2 \varphi) = 2\partial_2 \varphi\text{.}$$
So $2\partial_2 \varphi$ vanishes as well and so $P_1, P_2, Q_1, Q_2$. The same is for $Q_1 = P_2 = 0$
cases and we are left with $$P_1 = P_2 = 0$$ or $$Q_1 = Q_2 = 0$$ for
$\varphi : R^2 \longrightarrow S^2$ being conformal. This proves the proposition.\\
Orientation preserving conformal maps $\varphi^b :S^2 \longrightarrow S^2$
 are necessarily algebraic, that means: $$\varphi = \frac{P(z)}{Q(z)}\text{,}$$ with
P, Q  arbitrary polynomials [8].
These correspond to solutions with positive winding number. Negative winding number
solutions are represented by antiholomorphic (orientation reversing) maps.
The winding number is given by the degree of the polynomial P(z). This can be explained as follows:
Let $w_0$ be a regular value, i.e. $| \frac{\partial_i \varphi}{\partial_j x} | \neq 0$. It follows, that
$$P(z) - w_0 Q(z) = 0$$ has n different solutions (n being the degree of P(z)). But this is by
definition the winding number.
Finally, we have to clarify, whether there could be other finite energy solutions than the spin waves.
The answer is negative as shown by G. Woo [2].
As a final remark, let us have a look at the stereographic projection $\pi$
$$w = \frac{\varphi^1}{1 -\varphi^3} + i\frac{\varphi^2}{1 -\varphi^3}$$ from two dimensional
sphere down to the complex plane. In the same way the tangent vectors $\partial_i \varphi^a$ are
on the unit sphere are projected down into $\partial_i w$ in the complex plane, i.e.,
$$\pi^* : d\varphi^a \longrightarrow dw\text{.}$$
Since $\pi$ is conformal, it preserves right angles, but it reverses orientation:
$$\epsilon_{abc} \varphi^b \partial_i \varphi^c \longrightarrow - i \partial_i w\text{,}$$ i.e.,
$$ \# d\varphi^a \longrightarrow - i dw\text{.}$$
From linearity, it follows:
$$\epsilon_{ij} \partial_j  \varphi^a \longrightarrow \epsilon_{ij} \partial_j w\text{,}$$ i.e.,
$$*d\varphi^a \longrightarrow *dw\text{.}$$
Putting all this together, we see that the double self duality equation is projected
down to the self dual equation $$*dw =-idw\text{.}$$
This equation is conformal invariant in two
dimensions (and 4). Complex analysis tells us, that any solution $dw$ gives automatically a
holomorphic function $w$ and vice verse.
There seems to be a contradiction to Derrick's scaling argument [9] in that we have stable finite
energy solutions. But this is only apparent, since the sigma model represents a loophole
in Derrick's argumentation in not having a potential term in the Lagrangian. Generally pure
scalar field theories have only soliton solutions in dim = 1.
Derrick's argumentation runs as follows:
Consider a pure scalar field theory in D dimensions with Lagrange density
$$L = -1/2 \partial_{\mu} \varphi^a \partial^{\mu} \varphi^a - U(\varphi^a)\text{.}$$
The corresponding static energy functional is given by
$$H(\varphi^a ) = 1/2 \int_{R^d} \partial_i \varphi^a \partial^i \varphi^a + \int_{R^D} U(\varphi^a )$$
$$= H_1 (\varphi ) + H_2 (\varphi)\text{.}$$
Suppose $\varphi^a (x)$ is a static solution and consider the scaled configuration
$$\varphi^a_{\lambda} (x) = \varphi^a (\lambda x)\text{.}$$
The scaled configuration has static energy
$$H(\varphi^a_{\lambda} ) = 1/2 \int \lambda^2 \partial_i \varphi^a (\lambda x) \partial^i \varphi^a (\lambda x)d^D x$$ $$ + \int U(\varphi^a (\lambda x) d^D x$$
$$= 1/2 \int \lambda^{2-D} \partial_i \varphi^a (y) \partial^i \varphi^a (y) d^D y $$ $$+ \lambda^{-D} \int U(\varphi^a (y)) d^D y$$
$$= \lambda^{2-D} H_1 + \lambda^{-D} H_2\text{.}$$
If $\varphi^a$ is to be stable, $H$ must be stationary against variations of $\lambda$:
$$0 = \partial_{\lambda} H_{\lambda = 1} = (2 - D)H_1 - DH_2$$which implies $$(2-D)H_1 [\varphi ] = DH_2 [\varphi ]$$If D $> $ 2 the coefficients (2-D) and D have opposite sign. But since $H_1$ and $H_2$ are non-negative, they both must vanish.
This implies: A nontrivial static solution in a pure scalar field theory is unstable, if
the space dimension exceeds 2.
For dimension two we have $H_2 (\varphi^a ) = 0$, which can be fulfilled, if the potential
energy term is simply absent. This is the case in the two dimensional non-linear sigma model.
%%%%%%%%%%%%%%%%%%%%%%%%%%%%%%%%%%%%%%%%%%%%%%%%%%%%%%%%%%%%%%
\section{Vacuum structure}
A field theory which satisfies the conditions of Lorentz (Euclidean) invariance,
spectrum and locality, the vacuum is unique if and only if the n-point functions have
the cluster decomposition property [10]. Y. Iwasaki shows [4] that the correlation
functions have the cluster property for instanton or anti-instanton contributions but not
for instanton - anti-instanton contributions.
Similar ideas concerning the cluster property and topological non-trivial solutions [12],
instantons, apply to Yang-Mills theory [13].
\section{$\theta$ - vacuum and cluster property in QCD}
The $\theta$ vacuum is defined as the sum over all vacua with
topological winding number n, n $\epsilon \cal{Z}$,$|n>$ $$\theta >
= \sum e^{in\theta} |n > $$ The topological winding number or topological charge is defined as the second Chern class
or Pontryagin number $$ n = \frac{1}{8\pi^2} \oint Tr (FF^* ) d^4
x$$ $$= - \frac{1}{16\pi^2} \int \epsilon_{\mu \nu \rho \sigma}
Tr(A^{\nu} (\partial^{\rho} A^{\sigma} + \frac{2}{3} A^{\rho}
A^{\sigma} )$$ where F is the (non)- abelian fieldstrength tensor
and $F^* $ its Hodge dual $F^* = \frac{1}{2} \epsilon_{\mu \nu \rho
\sigma } F^{\rho \sigma } $ and Tr runs over space-time and color indices. The integral $\int $
runs over the boundary of space-time. Since we tackle Yang-Mils
theory, the vacua will be represented as pure gauge configurations
$$ A_{\mu} = g^{-1} \partial_{\mu} g , g \epsilon SU(2)$$ But the relevant point is that g will be a representative
of equivalence classes of $\pi_3  (S^3 )$ where $S^3$ will be the
boundary about which the integral $\oint$ of the topological charge
runs. g is a local gauge transformatio and hence amapping from
space-time to the parameter space of the gauge group (here SU(2) )
This is for SU(2) a three sphere $S^3$. We gain a mapping from
$S^3$ to $S^3$ by imposing aboundary condition at infinity $$g
\rightarrow  I$$Since g reaches the same value in all directios for the radius r, the space can be compactified by
a onepoint compactification $$\cal{R} \rightarrow S^3 $$ and we
have a mapping  $$g: S^3 \rightarrow S^3$$ Hence all pure gauge
potentials (vacua ) are divided in sectors, labeled by an integer
n, n being the second chern class or Pontryagin number, defined
above. We shall now discuss the vacuum structure and for this the
transition amplitude $$<n+Q| e^{-\tau H } |n> $$and we will see,
that in the presence of massless fermions ( $\tau $ is the
Euclidean time.) In the presence of fermions the action is $$S_E =
S_{YM} + S_{\Psi}$$ where $S_{\Psi} = i\int \overline{\Psi}
\gamma_{\mu}
\Psi$ with $D_{\mu} = i(\partial_{\mu} + A_{\mu}) \gamma_{\mu}$ $$lim_{\tau\to\infty}
<n+Q | e^{-H\tau } |n> $$ $$ = \int \cal{D} (A_{\mu} ) \int \cal{D}
(\psi ) \exp^{-S_E} $$ $$= \int \cal{D} (A_{\mu Q} ) \exp^{S_A} \det (\cal{D} (A_{\mu} ) ) $$
where $\cal{D} (A_{\mu Q} ) $ is the Dirac operator in the presence
of an external field $A_{\mu Q} $ . If the $A_{\mu Q} $ have non
trivial topological boundary conditions, as we have for the
instanton, this operator has zero eigenvalues according to the
Atiyah-Singer index theorem. Hence the integral above vanishes and
so the transition amplitude. The same result can also be gained by
using the ABJ - anomaly, but we refer to Coleman´ s lecture notes
(Erice, 1977 ) about instantons. At first sight, this seems to
suggest, that the $\theta$ vacuum is obsolete, but Callan showed,
that this must not be so. The situation is now, that, in the
presence of massless fermions, the topological vacua $|n>$ are
\begin{enumerate}
\item  the lowest energy eigenstates in the corresponding homotopy
sectors
\item  energetically are mutually degenerate, since the large
gauge transformations ( transformations altering the Chern class n
by a natural number.) commute with the Hamiltonian H.
\item  do not
tunnel into each other in the presence of massless fermions.
\end{enumerate}
Nevertheless Callan et al (1976) showed, that the  $\theta $vacuum
needs to be used as the correct vacuum for the sake of the cluster
property. Let A(x), B(x) be two local operators. Then cluster
decomposition property means $$<vac | A(x) B(y) |vac >
\rightarrow_{|x-y|\to\infty }$$ $$ <vac|A(x)|vac><vac|B(y)|vac> $$
Callan showed, that if we take only |n> vacua, cluster property
will be violated for operators of nonzero chirallity:Let B(x) be an
operator of chirality d $$[\overline{Q}_5 , B(x) ]_- = d B(x) $$
where $$\overline{Q}_5 = \int \overline{j}_{0,5} d^3x$$ $$ \int d^3 x
(j_{0,5} + \frac{1}{8\pi^2 } \epsilon_{\mu \nu \rho \sigma
}Tr [A^{\nu} (\partial^{\rho} A^{\sigma} + \frac{2}{3} A^{\rho}
A^{\sigma}) ]) $$ where Tr runs over the color indices. $J_{\mu 5}$
is defined by point splitting regularization method, since a
product of operators at one space point is ill defined:
$$ j_{\mu 5} = \overline{\psi} (x-\frac{1}{2} \epsilon )\gamma_{\mu} \gamma_5 P [e^{-\int_{x-\frac{\epsilon}{2}}^{x+\frac{\epsilon}{2}}
A_{\mu} d x^{\mu }} ] \psi (x+\frac{1}{2} \epsilon ) $$
where $\psi$ are spinor fields of the massless Dirac equations. At
first sight the chiral current $\overline{\psi} \gamma_5
\gamma_{\mu} \psi$ seems to be conserved, because the Dirac field is massless. But this
is only apparent, because, if we take $\epsilon \rightarrow 0$, we
find for the regularized current $j_{\mu 5} $ $$\partial_{\mu}
j_{\mu 5}  = \frac{i}{8\pi^2} Tr (FF^* ) $$ Apparently $j_{\mu 5}$
is no more conserved. But we can define a new one, which is
conserved (but gauge variant) $$ \overline{j}_{\mu 5} = j_{\mu 5} +
\frac{i}{4\pi^2} \epsilon_{\mu \nu \rho \sigma}
Tr [A^{\nu} (\partial^{\rho} A^{\sigma} + \frac{2}{3} A^{\rho }
A^{\sigma} )] $$ The charge $\overline{Q}_5 = \int d^3x
\overline{j}_{05} $ is not gauge invaiant but alters its values
under large gauge transformations by integer values. Let B(x) be a
local operator with nonzero chirality between topological
nontrivial vacua $|n>$ and $|m>$. We have $$[\overline{Q}_5 , H ]_-
= 0$$ because $\overline{Q}_5$ is a constant of time. Hence
$$2(m-n) <m|B(x)|n> = 0$$. Therefore $$<m|B(x)|n> = \delta_{d,
m-n}$$ But cluster decomposizion requires that$$\lim_{|x-y|
\to\infty} <n|B^{\dagger}(x) |n><n|B(y) |n> = 0$$But this is not true,
because, if we expand in intermediate states, we have nonzero
contributions $$<n|B^{\dagger} (x) |n+c><n+c|B(x)| n>$$ By
translation invariance we have$$<n|B^{\dagger} (0)
|n+c><n+c|B(0)|n>$$which will remain nonzero for $|x-y| \rightarrow
\infty$ violating cluster property.
\section{$\theta$ -  vacuum and the nonlinear sigma model}
The discussion of the violation of the cluster decompoaition
property ran by using operators of nonzero chirality. This seems
not to work with the $\sigma$ model, since, at first sight, we have
no fermionic degrees of freedom. But this is only apparent, since
we have a twodimensional QFT. But , since the calculations of Iwazaki 
are to lenghty, I will only cite the results and refer to the refence.
(The Structure of the Vacuum I)
Iwazaki finds a cluster property violation for instanton-antiinstanton 
calculations. But this means that the ground state of the O(3) model is 
degenerate and belong to inequivalent representations. From this he 
concludes, that a new kind of symmetry breaking occurs, which is not in 
conflict with Coleman's theorem, because the continuous O(3) symmetry is 
not broken.

%%%%%%%%%%%%%%%%%%%%%%%%%%%%%%%%%%%%%%%%%%%%%%%%%%%%%%%%%%%%%%
%
% Bibliography
%
%%%%%%%%%%%%%%%%%%%%%%%%%%%%%%%%%%%%%%%%%%%%%%%%%%%%%%%%%%%%%%

\end{document}